\documentclass[aps,prb,twocolumn,showpacs,superscriptaddress,floatfix,nofootinbib]{revtex4}

\usepackage{amsfonts}
\usepackage{amsmath}
\usepackage{amssymb}
\usepackage{graphicx}
\usepackage{times}
\usepackage{subfigure}
\usepackage{array}
\usepackage{hyperref}

\hypersetup{colorlinks=true}

\begin{document}

\title{Competition between three-sublattice order and superfluidity in the quantum 3-state Potts model of ultracold bosons and fermions on a square  optical lattice}

\author{Laura Messio}
\affiliation{Institut de Physique Th\'eorique (IPhT), CEA, CNRS, URA 2306, F-91191 Gif-sur-Yvette, France}

\author{Philippe Corboz}
\affiliation{Theoretische Physik, ETH Zurich, 8093 Zurich, Switzerland}

\author{Fr\'ed\'eric Mila}
\affiliation{Institute of Theoretical Physics, \'Ecole Polytechnique F\'ed\'erale de Lausanne (EPFL), CH-1015 Lausanne, Switzerland}

\date{\today}

\begin{abstract}
We study a quantum version of the three-state Potts model that includes as special cases the effective models of bosons and fermions on the square lattice in the Mott insulating limit. It can be viewed as a model of quantum permutations with amplitudes $J_\parallel$ and $J_\perp$ for identical and different colors, respectively. For $J_\parallel=J_\perp>0$ it is equivalent to the  SU(3) Heisenberg model, which describes the Mott insulating phase of 3-color fermions, while the parameter range $J_\perp<\min(0,-J_\parallel)$ can be realized in the Mott insulating phase of 3-color bosonic atoms.  
Using linear flavor wave theory, infinite projected entangled-pair states (iPEPS), and continuous-time quantum Monte-Carlo simulations, we construct the full $T=0$ phase diagram, and we explore the $T\neq0$ properties for  $J_\perp<0$. 
For dominant antiferromagnetic $J_\parallel$ interactions, a three-sublattice long-range ordered stripe state is selected out of the ground state manifold of the antiferromagnetic Potts model by quantum fluctuations. 
Upon increasing  $|J_\perp|$, this state is replaced by a uniform superfluid for $J_\perp<0$, and by an exotic three-sublattice superfluid followed by a two-sublattice superfluid for $J_\perp>0$. 
The transition out of the uniform superfluid (that can be realized with bosons) is shown to be a peculiar type of Kosterlitz-Thouless transition with three types of elementary vortices. 
 \end{abstract}

\maketitle

\section{Introduction}
\label{sec:introduction}
The possibility to realize N-color Fermi- or Bose-Hubbard models with ultra-cold atoms in optical lattices has attracted increasing interest recently.~\cite{wu2003,honerkamp2004,cazalilla2009,fukuhara09, gorshkov2010,taie10,sugawa11} With fermionic alkaline-earth atoms (and with the alkaline-earth like atomYb), it is possible to realize systems with up to N=10 different flavors (or colors) of fermions by exploiting different nuclear spin states,~\cite{gorshkov2010} while 3-color bosonic systems can be created with bosonic alkali atoms with a hyperfine spin $F=1$, e.g. $^{23}$Na, $^{39}$K, and $^{87}$Rb. A good starting point to describe these systems is  
the SU(N) Hubbard model  given by
  \begin{eqnarray}
  H=&-&t\sum_{\langle ij \rangle,\alpha} (\hat c^\dagger_{i,\alpha}\hat c^{\vphantom{\dagger}}_{j,\alpha}+H.c.) \nonumber \\ 
  &+&\sum_{i,\alpha<\beta}U_\perp \hat n_{\alpha,i} \hat n_{\beta,i}
  +\sum_{i,\alpha}\frac{U_\parallel}2 \hat n_{\alpha,i} (\hat n_{\alpha,i}-1),
  \label{eq:Ham_Hubbard}
  \end{eqnarray}
where $\langle ij\rangle$ denotes a sum over all pairs of nearest neighbors, and  the operators $\hat c^\dagger_{i,\alpha}$ and $\hat c^{\vphantom{\dagger}}_{i,\alpha}$ create and annihilate a bosonic or fermionic particle with color $\alpha$ on site $i$, respectively. The second term describes an on-site interaction with amplitude $U_\perp$ between particles with different colors $\alpha \neq \beta$, with $\hat n_{\alpha,i}=\hat c^\dagger_{i,\alpha}\hat c^{\vphantom{\dagger}}_{i,\alpha}$,  whereas the third term corresponds to the on-site interaction between particles with the same color. For fermions, the last term vanishes because of the Pauli exclusion principle.  
For femionic alkaline-earth atoms, this description is very accurate because, since the interactions between the particles do not depend on the nuclear spin, these systems exhibit a full SU(N) symmetry. In the bosonic case, there is in general an additional on-site interaction
proportional to the square of the total spin that can be antiferromagnetic (as in $^{23}$Na) or ferromagnetic (as in $^{87}$Rb).~\cite{law}
This interaction is small - it vanishes if the scattering lengths of the spin-0 and spin-2 channels are equal - and 
we will not consider it in the present paper. 

At filling $1/N$ (one particle per site) and for large enough repulsion the system is in a Mott insulating phase, and color fluctuations can be described by the effective Hamiltonian:
\begin{equation}
\widehat H= J_\parallel\sum_{\alpha\langle ij\rangle} \widehat S_i^{\alpha\alpha}\widehat S_j^{\alpha\alpha}
+J_\perp\sum_{\alpha\neq\beta\langle ij\rangle}\widehat S_i^{\alpha\beta}\widehat S_j^{\beta\alpha},
\label{eq:Ham}
\end{equation}
where $\widehat S^{\alpha\beta}$ is defined by $\widehat S^{\alpha\beta}|\gamma\rangle=\delta_{\beta\gamma}|\alpha\rangle$.
The values of the coupling constants $J_\parallel$ and $J_\perp$ depend on the statistics of the particles. To second order in $t/U_\parallel$ and $t/U_\perp$, they are given by $J_\parallel=J_\perp=2t^2/U_\perp$ for fermions, and by $J_\parallel=2t^2(1/U_\perp-1/U_\parallel)$ and $J_\perp=-2t^2/U_\perp$ for bosons.~\cite{kuklov,altman} Thus,  the experimentally accessible parameter range is given by $J_\parallel=J_\perp>0$ for fermions, and $J_\perp<\min(0,-J_\parallel)$ for bosons. Throughout the paper we will use the notation $J_\parallel=\cos\theta$ and $J_\perp=\sin\theta$ to parametrize the Hamiltonian by a single parameter $\theta$.

In the fermionic case with equal couplings $J_\parallel=J_\perp$ ($\theta=\pi/4$) the Hamiltonian corresponds to the antiferromagnetic (AF) SU($N$) symmetric Heisenberg model. 
For $N=2$, we recover the well-known AF Heisenberg $S=1/2$ model. The case of larger $N$ has been the subject of many theoretical studies, in which a large variety of different ground states have been found. For example, for $N=3$,  a three sublattice N\'eel ordered state has been predicted on the square and triangular lattice,\cite{SU3_Toth,SU3_Bauer} whereas on the honeycomb and the kagome lattice the ground state can be understood as a generalized valence-bond solid state.~\cite{arovas2008,hermele2011,Corboz12_simplex,Zhao12,Corboz13_su3hc}  Even more exotic ground states have been predicted for larger $N$, such as a dimer-N\'eel ordered state on the square lattice,~\cite{corboz11-su4} an algebraic spin-orbital liquid on the honeycomb lattice~\cite{Corboz12_su4} (for $N=4$), and chiral spin liquids.~\cite{hermele2009,hermele2011,szirmaiG2011}
  
The anisotropic case of the model of Eq.~\eqref{eq:Ham} ($J_\parallel \neq J_\perp$)  can be seen as a generalization of the anisotropic $S=1/2$ Heisenberg (XXZ) model for $N=2$, for which the phase diagram on the square lattice is well known: if $|J_\parallel| > |J_\perp|$ the diagonal terms in the Hamiltonian are dominant and the ground state is an antiferromagnetic (ferromagnetic) state for $J_\parallel>0$ ($J_\parallel<0$), whereas if $|J_\parallel| < |J_\perp|$ the off-diagonal terms are dominant, and the ground state is a superfluid with uniform (2-sublattice) superfluid order parameter for  $J_\perp<0$ ($J_\perp>0$). However, for larger $N$ the ground state phase diagram is unknown.

In this article we focus on the case $N=3$, and study the full ground state phase diagram of the model~\eqref{eq:Ham} on a square lattice, using linear flavor wave theory (LFWT, Sec.~\ref{sec:LFWT}), infinite projected entangled-pair states (iPEPS, Sec.~\ref{sec:iPEPS}) and continuous-time world-line quantum Monte-Carlo (CTQMC, Sec.~\ref{sec:QMC}). With the latter method we also explore the finite temperature phase diagram for  $J_\perp<0$, in which case there is no negative sign problem.

\section{Linear flavor wave theory (LFWT)}
  \label{sec:LFWT}

  \subsection{The method}
  \label{sec:LFWT_method}
  
    The LFWT is a generalization of the spin-wave expansion of $N=2$ models to $N>2$.\cite{LFWT}
    The large parameter of LFWT is the filling denoted by $M$ (it is usually called spin and denoted by $S$ for $N=2$). 
    The representation used on each site corresponds to a Young tableau with $M$ boxes arranged horizontally. 
    
    First, one chooses a \textit{classical} state $|\phi\rangle$ to which quantum fluctuations will be added. 
    These states (the product states, described in Sec.~\ref{sec:product_states}) are the ground states of the $M=\infty$ model and are the subject of the next subsection. 
    We then perform a local SU($N$) transformation such that the image state $|\phi'\rangle$ has $M$ bosons of color $\alpha=N$ on each site. 
    Next the transformed Hamiltonian (expressed in terms of new operators) is expanded in powers of $1/\sqrt M$ and its truncation to quadratic operators is solved by a Bogoliubov transformation. 
    The results for our model are given in Sec.~\ref{sec:LFWT_results}. 
    The possible phase transitions are discussed in Sec.~\ref{sec:transitions}.

  \subsection{Product states}
  \label{sec:product_states}
  
    Let us define $\widehat a_{i\alpha}^\dag$ as an operator creating a bosonic particle of color $\alpha$ at site $i$ and $|0\rangle$ as the vacuum of particles. 
    A state $|\phi_i\rangle$ at site $i$ can be fully described by a unitary $N$ component complex vector $\mathbf z_i=(z_{i1},z_{i2},\dots,z_{iN})$ such that $|\phi_i\rangle=\sum_\alpha z_{i\alpha}\widehat a_{i\alpha}^\dag|0\rangle_i$. 
    The local U(1) phase of the $\mathbf z_i$ vector has no importance.
    Thus the state corresponds to a single point of the complex projective space CP(2), but the vectorial notation will be conserved to keep notations simple. 
    
    Product states are simply tensor products of single site states, i.e. non-entangled states. 
    They are the generalization of the classical spin states to the $N$ color models (for $N=2$, product states are fully magnetized states characterized by a unitary three-dimensional real vector at each site). 
    For a lattice of $N_s$ sites, a product state is described by $N_s(N-1)$ complex numbers, to be compared with $(N^{N_s}-1)$ for any state of the full Hilbert space. 
    In this reduced space, the minimal energy is variational. 
    Since the square lattice with nearest-neighbor interactions is bipartite, 
    the lowest energy state minimizes the energy of each link $\langle ij\rangle$ which depends only on $\mathbf z_i$ and $\mathbf z_j$:
    \begin{equation}
    E_{ij}=\left(J_\parallel-J_\perp\right)\sum_\alpha\left|z_{i\alpha}^*z_{j\alpha}\right|^2 +J_\perp\big|\sum_\alpha z_{i\alpha}^*z_{j\alpha}\big|^2.
    \end{equation}
    
\begin{figure}
      \begin{center}
        \includegraphics[width=0.35\textwidth]{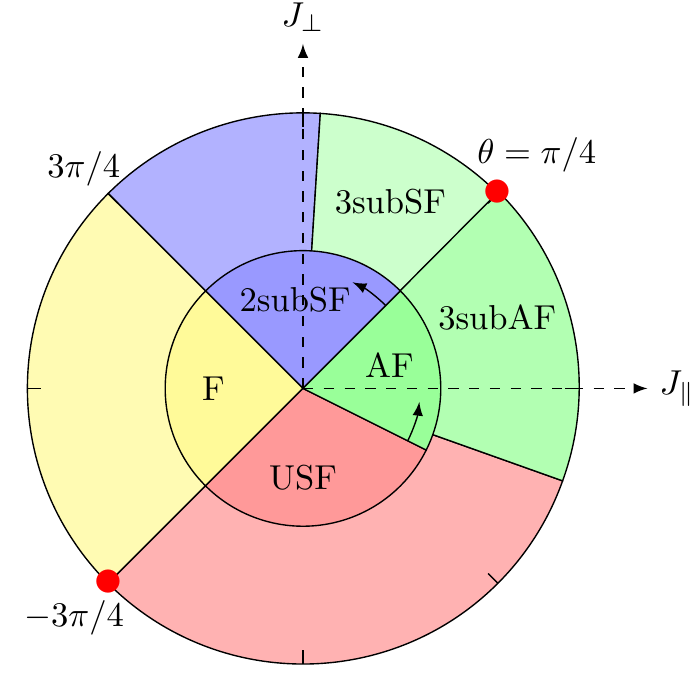}
      \end{center}
      \caption{Phase diagram obtained by linear flavor wave theory (inner circle) and iPEPS (outer circle). 
      The different phases, 2-sublattice superfluid phase (2subSF), 3-sublattice antiferromagnetic phase (3subAF), 3-sublattice superfluid phase (3subSF), uniform superfluid (USF), and ferromagnetic (F), are described in the text. 
      The two arrows indicate the direction of the displacement of the boundaries when harmonic fluctuations are taken into account. 
      $(J_\parallel,\,J_\perp)=(\cos\theta,\,\sin\theta)$. 
      \label{fig:product_states}}
    \end{figure}

    Depending on $\theta$, there are four phases of product ground states for $N=3$ (see the inner circle of Fig.~\ref{fig:product_states}):

    \begin{itemize}

    \item the 2-sublattice superfluid (2subSF) phase ($\pi/4<\theta<3\pi/4$). 
    There are three distinct ground state manifolds. 
    The states of one of them are characterized by a phase $\phi$ such that the vectors on the two sublattices are $\left(\frac{e^{i\phi}}{\sqrt2},\frac{e^{-i\phi}}{\sqrt2},0\right)$ and $\left(-\frac{e^{i\phi}}{\sqrt2},\frac{e^{-i\phi}}{\sqrt2},0\right)$. 
    The two remaining manifolds are obtained by a cyclic permutation of the vector components. 
    This leads to a set of ground states homeomorph to $C_3\times U(1)$. 
    The ground state energy per bond is 
    $E_{\rm 2subSF}=(J_\parallel-J_\perp)/2$.
    
    \item the antiferromagnetic (AF) phase ($\arctan(-1/2)<\theta<\pi/4$): 
    the vectors are chosen among $(1,0,0)$, $(0,1,0)$ and $(0,0,1)$ and set on the lattice in such a way that two neighboring sites have different vectors. 
    These states are the ground states of the classical antiferromagnetic Potts model, which has an infinite degeneracy, and a ground state energy per bond of 
    $E_{\rm AF}=0$.
    
    \item the uniform superfluid (USF) phase ($-3\pi/4<\theta<\arctan(2)-\pi/2$): all vectors are identical on the lattice and the components fulfill $|z_{i\alpha}|=1/\sqrt3$.
    The ground state manifold is homeomorph to a torus ($U(1)\times U(1)$). 
    The states can be parametrized by $\mathbf z=(\frac{e^{i\alpha}}{\sqrt3},\frac{e^{i\psi}}{\sqrt3},\frac{e^{i\phi}}{\sqrt3})$ with $\alpha+\psi+\phi=0$, with an energy
     $E_{\rm USF}=(2J_\perp+J_\parallel)/3$ per bond.
     
    \item the ferromagnetic (F) phase ($-5\pi/4<\theta<-3\pi/4$): all vectors are identical on the lattice and equal to  either $(1,0,0)$, $(0,1,0)$ or $(0,0,1)$. 
    These states are the ground states of the classical ferromagnetic Potts model. 
    In fact, these states are the exact quantum ground states in this range of $\theta$. 
    The ground state manifold is homeomorph to  $C_3$, and the energy is  $E_{F}=J_\parallel$ per bond.
   
    \end{itemize}
    
    Two values of $\theta$ do not break the SU(3) symmetry (red points in Fig.~\ref{fig:product_states}):
    
    \begin{itemize}
    \item $\theta=\pi/4$: This is the AF SU(3) symmetric point. 
    Every state with perpendicular neighboring vectors is a ground state and there exist an extensive number of such states. 
    At this point, the 2 sublattice AF and the 2subSF phases of Fig.~\ref{fig:product_states} are linked by a global SU(3) transformation. 
    \item $\theta=-3\pi/4$: This is the F SU(3) symmetric point. 
    The ground states are the ferromagnetic states, i.e. those with uniform $\mathbf z_{i}$. 
    The ground state manifold is homeomorph to  $CP(2)$. 
    At this point, the F and the USF phases of Fig.~\ref{fig:product_states} belong to this manifold, and they are linked by a global SU(3) transformation. 
    \end{itemize}
    
    For $N=2$, since the square lattice is bipartite, we have the same properties for $\theta$ as for $-\theta$  and this change of parameters can be compensated by a multiplication of the vectors of one sublattice by the matrix $i\sigma_z$ (rotation of the spins of $\pi$ around the $z$ axis in the corresponding spin model). 
    It is this property that solves the sign problem of the Heisenberg spin model on the square lattice. 
    Then $\theta=3\pi/4$ and $-\pi/4$ are too SU(2) symmetric points for $N=2$.
    For $N=3$ we have a reduced SU(2) symmetry at these angles: states with one vector component equal to zero everywhere and their transforms have the same energy for $\theta$ and $-\theta$, thus the subset of such ground states of a SU(3) symmetric point are sent to degenerate subsets for opposite $\theta$.
    It gives a continuum of ground states for $\theta=3\pi/4$ including both the 2subSF and F states. 
    At $\theta=-\pi/4$, we do not have such a continuum since the $\mathbf z$ vectors of the ground states have three non zero components. 

  \subsection{Quantum fluctuations at the harmonic level}
  \label{sec:LFWT_results}
    Harmonic quantum fluctuations lift some of the degeneracies found in the product state analysis of the previous section, however, not all of them: when degenerated ground states are connected via a symmetry of the Hamiltonian, then the harmonic (and higher order) fluctuations cannot lift their degeneracy. 
    This is the case at the SU(3) symmetric point of the F-USF boundary ($\theta=-3\pi/4$), because the two product states can be transformed into each other by a global SU(3) transformation. 
    
    The degeneracy is not lifted either at the F-2subSF phase boundary ($\theta=3\pi/4$), but for a different reason:
    the F and 2subSF states both belong to a continuum of states linked by a set of transformations homeomorph to SU(2), but that are not in the group of the Hamiltonian symmetries (the origin of this reduced symmetry was given in the previous section). 
    The states with vectors $(a,b,0)$ and $(-a,b,0)$ on both sublattices have the same energy for any complex numbers $a$ and $b$ such that $|a|^2+|b|^2=1$. 
    The ground state manifold is thus extended to SU(2) and contains both the F and the 2subSF state. Since such states are exact quantum ground states (they have the same energy per bond as an isolated bond), the degeneracy remains at any order. 
    
    At all other boundaries of the LFWT phase diagram and for any value of $\theta$ in the AF phase (since there is an extensive degeneracy of product-states), the term of order $M$ will lift the classical degeneracy. 
    To calculate the first order quantum correction, we first apply a local SU($N$) transformation to the operators $\widehat S_i^{\alpha\beta}$ that become 
    ${{\widehat S}^{'\alpha\beta}_i}$ such that the vector $\mathbf z_i$ is transformed into $(0,0,1)$. 
    We then perform the following replacement by bosonic operators in the modified Hamiltonian:
    \begin{eqnarray}
        \widehat S'^{\beta\alpha}&=&\widehat b_\beta^\dagger \widehat b_\alpha\nonumber\\
        \widehat S'^{\alpha N} &=&\widehat b_\alpha^\dagger \sqrt{M}+O(M^{-\frac12})\\
        \widehat S'^{NN} &=& M -\sum_{\beta\neq N} \widehat b_\beta^\dagger \widehat b_\beta\nonumber, 
    \end{eqnarray}
    ($\alpha$ and $\beta$ are colors different from $N$) and solve the Hamiltonian on the square lattice neglecting the $O(\sqrt M)$ terms. 
    
\begin{figure}
      \begin{center}
        \includegraphics[width=0.35\textwidth]{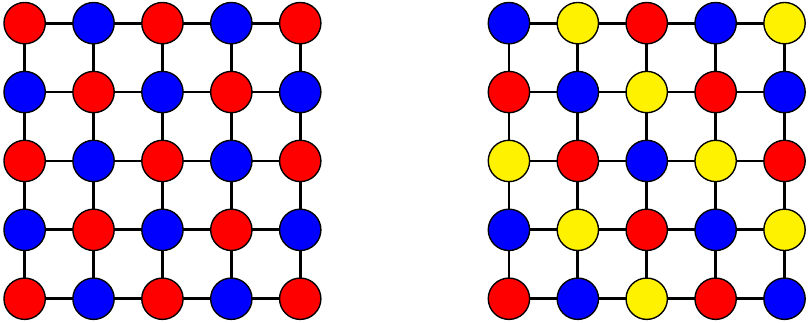}
      \end{center}
      \caption{The two states considered in the calculation of the first non trivial order of LFWT among the infinite number of AF product states: 2subAF and 3subAF.
      \label{fig:23}}
    \end{figure}

    Among the extensive number of AF states, we will only consider the 2 sublattice order (noted 2subAF) and the 3 sublattice stripe order\cite{SU3_Bauer} (noted 3subAF) depicted in Fig.~\ref{fig:23}. 
    Each state of Sec.~\ref{sec:product_states} is stable up to the first order correction in the interval of $\theta$ where is is the ground state. 
    Moreover, the $2$-$3$subAF are both stable until $\theta=-\pi/4$. 
    For a product state, stability means that  the derived quadratic Hamiltonian has a ground state (the Bogoliubov transformation is possible). 
    At the classical boundary between two orders, we can compare the corrections of order $M$ of both phases and determine which one has the lowest energy to this order. 
    Since simply taking $M=1$ does not give an accurate estimate of the energy (one would need further orders), one cannot be quantitative regarding the boundary, and we just indicate in which direction it shifts due to harmonic fluctuations (arrows on Fig.~\ref{fig:product_states}). 
    
    The corrections for the 2subAF and 2subSF phases are the same at the AF SU(3) symmetric point because of the symmetries, but since 2subAF is degenerate with other product states with different symmetry properties, the first order corrections of these other states will be different and possibly more favorable. 
    Thus, the boundary between AF and 2subSF phases can still be pushed towards larger $\theta$'s. 
    
    We now give the numerical results and the evolutions for all boundaries. 
    At the AF-USF boundary, the correction per site is $-0.12M$ for the 3subAF, $-0.058M$ for the 2subAF and $-0.28M$ for the USF phase. 
    Thus we expect that the USF phase is stable until $\theta>\arctan(2)-\pi/2$. 
    
    At the AF-2subSF boundary, the correction is $-0.51M$ for the 3subAF and $-0.22M$ for the 2subAF and 2subSF. 
    It suggests that the 3subAF survives for $\theta>\pi/4$. 
    But the degenerate 3subAF states obtained by global SU(3) transformations at the SU(3) symmetric point do not have the same classical energy as soon as we are not exactly at the $\theta=\pi/4$ point. 
    For smaller angles, the better states are those described previously, with vectors among $(1,0,0)$, $(0,1,0)$ and $(0,0,1)$. 
    For larger angles, we can reach $E=(J_\parallel-J_\perp)/3$ on a link with vectors among 
    $(\frac{e^{i \alpha}}{\sqrt3},\frac{ e^{i \psi}}{\sqrt3},\frac{e^{i \phi}}{\sqrt3})$, 
    $(\frac{e^{i \alpha}}{\sqrt3},j\frac{ e^{i\psi}}{\sqrt3},j^2\frac{e^{i\phi}}{\sqrt3})$ and 
    $(\frac{e^{i \alpha}}{\sqrt3},j^2\frac{ e^{i\psi}}{\sqrt3},j\frac{e^{i\phi}}{\sqrt3})$, with $j=e^{2i\pi/3}$ and $\alpha+\psi+\phi=0$. 
    It will thus be this 3 sublattice superfluid stripe order dressed by quantum fluctuations that will survive for $\theta>\pi/4$, noted 3subSF in Fig.~\ref{fig:product_states}
    (the ground state manifold is homeomorph to ${\mathbb Z_2}^2\times U(1)^2$). 
    
    All these predictions will be confirmed by the iPEPS and/or QMC measurements. 
    
  \subsection{Symmetries and phase transitions}
  \label{sec:transitions}
  
    From the broken symmetries of the product-states of Sec.~\ref{sec:product_states}, we can have an idea of the finite temperature phase transitions. 
    Equivalently, we can look at the topology of the ground state manifold. 
    The Mermin-Wagner theorem\cite{MerminWagner} states that a connected ground state manifold cannot give rise to a finite order phase transition in two dimensions, except at zero temperature. 
    This is the case for the USF phase. 

    When the ground state manifold consists of several disconnected regions (when some points cannot be linked by a continuous path in the manifold), then there is at least one first or second order phase transition. 
    The F and 2subSF phases will give rise to such a phase transition with a $C_3$ order parameter. 
    The 3subAF and 3subSF phases break a $P_3\times \mathbb Z_2$ symmetry, that can be restored through one or more transitions ($P_3$ is the group of permutations for three elements). 
    Moreover, it is not excluded that 2subAF becomes more stable than 3subAF at finite temperature,\cite{SU3_Toth} and that it gives rise to its own transitions. 
    
    Besides the finite order phase transitions, infinite order phase transitions associated to topological defects\cite{Mermin} can occur when the components of the ground state manifold are not simply connected (some paths are not continuously shrinkable to a point). 
    This is the case for the USF, 2subSF, and 3subSF phases, possessing pairs of defects at low temperature. 
    The defects encountered in the 2subSF phase are in $\mathbb Z$ (the first homotopy group of $U(1)$ is $\pi_1(U(1))=\mathbb Z$), whereas those of the USF and 3subSF phases are in $\pi_1(U(1)\times U(1))=\mathbb Z\times \mathbb Z$.
    Thus, a Kosterlitz-Thouless (KT) transition\cite{KosterlitzThouless} could occur in the 2subSF phase and a KT type phase transition in the USF and 3subSF phases. 
    The defects encountered in these two phases are the same as the dislocations of triangular two dimensional crystals and the phase transition is similar to its melting. \cite{Melting_Nelson} 
    
    The case of the USF phase is detailed below. 
    Among the three phases with defects, USF is the only one in which defects do not coexist with a discrete symmetry breaking and for which the occurence of a topological phase transition makes no doubt. 
    In the two other phases, 2subSF and 3subSF, we have both defects and discrete symmetry breaking. 
    It can lead to a variety of phase transitions, with the possibility of fractional vortices, as in the case of the AF XY model on the triangular lattice. \cite{KorshunovI, KorshunovIII, FFXY_numerics}
    The larger discrete symmetry and the more complicated topological defects for the 3subSF phase require further investigations about the nature of the phase transitions, which go beyond the scope of the present paper.  

    We concentrate here on the USF topological phase transition. 
    If such a transition occurs in the 3subSF phase, the same description can be easily adapted. 
    We neglect the $\mathbf z$ component fluctuations and construct a subset of product states interacting via the same Hamiltonian and possessing the same superfluid properties.
    More precisely, we only consider product states whith $\mathbf z_j=\left(\frac{e^{i\alpha_j}}{\sqrt3},\frac{e^{i\psi_j}}{\sqrt3},\frac{e^{i\phi_j}}{\sqrt3}\right)$. 
    We can fix the gauge such that $\alpha_j+\psi_j+\phi_j=0$ and parametrize a the state of a site state with only two real 
    parameters $\eta_j=\alpha_j$ and $\mu_j=\frac{\psi_j-\phi_j}{\sqrt3}$. Since $J_\perp<0$, we note $J_\perp=-|J_\perp|$ so that the sign of the different quantities that will follow is obvious. 
    The energy on a bond then reads: 
    \begin{eqnarray}
    H_{ij}&=&-\frac{2|J_\perp|}9\left(
      \cos\left(-\sqrt3\Delta \mu\right)+\cos\left(\frac32\Delta \eta-\frac{\sqrt3\Delta \mu}2\right)\right.\nonumber\\    &&\left.+
      \cos\left(\frac32\Delta \eta-\frac{\sqrt3\Delta \mu}2\right)\right)+\frac{J_\parallel}3, 
    \end{eqnarray}
    where $\Delta \eta=\eta_j-\eta_i$ and $\Delta \mu=\mu_j-\mu_i$. 
    An expansion in $\Delta \eta$ and $\Delta \mu$ gives:
    \begin{equation}
    H_{ij}\simeq E_0+\frac{|J_\perp|}2(\Delta \eta^2+\Delta \mu^2), 
    \end{equation}
    with the energy of a uniform state $E_0=\frac{J_\parallel}3-\frac{2|J_\perp|}3$. 
    Note that a uniform state is labelled by a point $(\eta,\mu)$ of the order parameter space, whose topological properties will be discussed below. 
    Taking the continuum limit of this model on the square lattice, we obtain 
    \begin{equation}
    H=\int d\mathbf r \left(2E_0+\frac{|J_\perp|}2\left((\vec\nabla \eta)^2+(\vec\nabla \mu)^2\right)\right), 
    \label{eq:Ham_cont}
    \end{equation}
    If we look for the local minima of the energy, we obtain the two conditions 
    \begin{equation}
    \label{eq:local_min}
    \vec\nabla^2 \eta=0, \qquad  \vec\nabla^2 \mu=0.
    \end{equation}
    These conditions are verified if $\eta$ and $\mu$ are constant (global minimum of energy, $2E_0$ per surface unit), but also for configurations with local singularities called vortices (metastable states). 
    This model looks very similar to two uncoupled XY models $H_{ij}=-|J_\perp|(\cos(\Delta \eta)+\cos(\Delta \mu))$, that also leads to
    Eq.~\eqref{eq:Ham_cont} in the continuum limit, but they are different because of the nature of the topological defects. The topological defects are linked to the set of equivalent points in the covering group of the order parameter space: the plane $(\eta,\mu)$. 

    In the $XY$ model, the covering group of the order parameter space $U(1)$ is the real line $\mathbb R$, and two points are equivalent if their distance is a multiple of $2\pi$. 
    A closed path in the order parameter space corresponds to a path in $\mathbb R$ between two equivalent points, characterized by a relative integer $n_\eta$, the distance between these two points divided by $2\pi$, also called the winding number. 

    In the two uncoupled $XY$ models, the covering group of the order parameter space is the real plane $\mathbb R^2$, and equivalent points are on a square lattice of spacing $2\pi$. 
    A closed path in the order parameter space is now characterized by two relative integers $(n_\eta,n_\mu)$. 

\begin{figure}
    \begin{center}
      \includegraphics[width=0.5\textwidth]{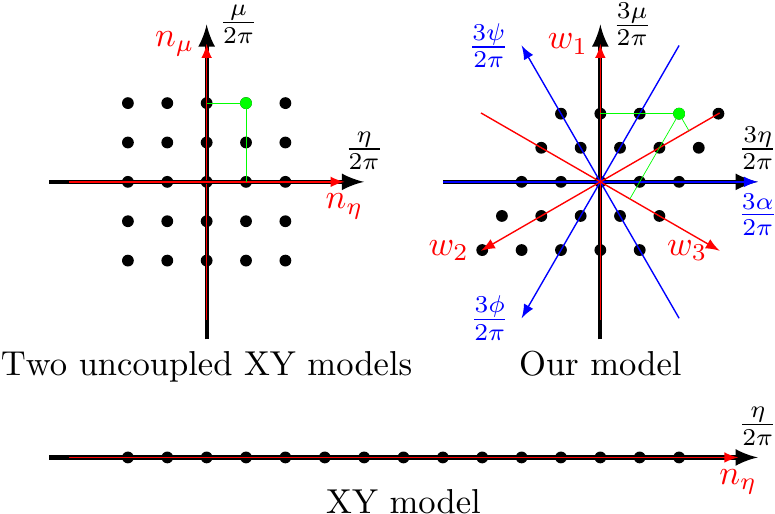}
    \end{center}
    \caption{\label{fig:covering}
    Covering groups of the order parameter space of different models. 
    All points equivalent to the origin are depicted. 
    The winding numbers of the green points are indicated by green lines. }
    \end{figure}

    In our model (two coupled $XY$ models), the covering group of the order parameter is still a plane, but equivalent points $(\eta,\mu)$ form a triangular lattice of spacing $2\pi$ (related to the Burger's vectors of a triangular crystal\cite{Melting_Nelson}). 
    The most symmetric way to characterize the topological properties of a closed path in the order parameter space is to use three relative integers $(w_1,w_2,w_3)$ defined by 
    \begin{eqnarray}
    w_1&=&\frac{1}{2\pi}\oint d\mathbf l\cdot \sqrt3\vec\nabla \mu(\mathbf r)\\
    w_2&=&\frac1{2\pi}\oint d\mathbf l\cdot \left(-\frac32\vec\nabla \eta(\mathbf r)-\frac{\sqrt3}{2}\vec\nabla \mu(\mathbf r)\right)\\
    w_3&=&\frac1{2\pi}\oint d\mathbf l\cdot \left(\frac32\vec\nabla \eta(\mathbf r)-\frac{\sqrt3}{2}\vec\nabla \mu(\mathbf r)\right), 
    \end{eqnarray}
    linked by $w_1+w_2+w_3=0$ and represented on Fig.~\ref{fig:covering}. 

    We now want to calculate the energy $E_v(w_1,w_2,w_3)$ of a configuration with a vortex of winding numbers $(w_1,w_2,w_3)$ at position $\mathbf r=0$ and satisfying Eq.~\eqref{eq:local_min}. 
    Since the problem has a spherical symmetry, $\eta(\mathbf r)$ and $\mu(\mathbf r)$ only depend on the modulus $r$. 
    Thus $|\vec\nabla \mu(\mathbf r)|=\frac{w_1}{\sqrt3 r}$ and $|\vec\nabla \eta(\mathbf r)|=\frac{w_3-w_2}{3r}$.
    We obtain from Eq.~\eqref{eq:Ham_cont}
    \begin{equation}
    E_v(w_1,w_2,w_3)=\frac{2|J_\perp|\pi}{9}(w_1^2+w_2^2+w_3^2)\ln\frac La .
    \end{equation}
    We needed two cut-off length: the lattice spacing $a$ and the lattice size $L$. 
    Six elementary vortices exist, with an energy $\frac{4\pi}{9}|J_\perp|\ln\frac La$: $(w_1,w_2,w_3)=(1,-1,0)$, $(0,1,-1)$, $(-1,0,1)$ and their antivortices with opposite $w$'s. 
    
    We have here detailed the case of a three color USF phase, but similar $N$ color USF phases exist. 
    Then, the number of elementary vortices would be $N-1$. 
    The transition associated to these vortices is numerically characterized through the winding number (see Sec.~\ref{sec:QMC}). 

\section{Infinite projected entangled-pair states}
\label{sec:iPEPS}
  \subsection{Method}
  \label{sec:iPEPS_method}

An infinite projected-entangled pair state (iPEPS) is a variational tensor network ansatz for two-dimensional ground state wave functions in the thermodynamic limit,\cite{verstraete2004, jordan2008} where the accuracy can be systematically controlled by the so-called bond dimension $D$. 
It consists of a unit cell of rank-5 tensors with one tensor per lattice site, which is periodically repeated on the lattice.~\cite{corboz2011} The first index of a tensor carries the local Hilbert space of the corresponding lattice site, and  the four remaining indices, called auxiliary bonds,  connect the tensor to its four nearest neighbors. Details on the iPEPS method  can be found in Refs.~\onlinecite{jordan2008, corboz2010, corboz2011}, and also in Ref.~\onlinecite{SU3_Bauer}, where we used the same approach to study the AF SU(3) symmetric point of the current model. 

For the experts we note that the optimization of the tensors is done through imaginary time evolution using the so-called simple update,\cite{vidal2003-1, orus2008, jiang2008, corboz2010} and we verified some simulations results also with the full update.\cite{corboz2010} 
The corner-transfer matrix method~\cite{nishino1996, orus2009-1,corboz2010} is used to  contract the tensor network, where the  error of the approximate contraction can be controlled by the boundary dimension $\chi$. 
The error due to the finite $\chi$ is small compared to the symbol sizes. 
To simulate the state with 3-sublattice (diagonal) order we used tensors with $\mathbb{Z}_q$ symmetry~\cite{singh2010,bauer2011} to improve the efficiency. In the superfluid phases we cannot exploit this symmetry since it is spontaneously broken in the thermodynamic limit.

By performing simulations with different (rectangular) unit cell sizes in iPEPS we can represent states with different type of lattice symmetry breaking. For example, a two-sublattice  state requires a $2\times 2$ unit cell with two different tensors $A$ and $B$ arranged in a checkerboard order, whereas a three-sublattice state requires a $3\times3$ unit cell with three tensors arranged in  a stripe order. In practice we perform simulations with different unit cell sizes and check which ansatz yields the lowest variational energy. Finally we note that the optimization scheme (imaginary time evolution)  requires an ansatz with at least two different tensors $A$ and $B$, and this is why the smallest unit cell we consider is $2\times2$ (containing two different tensors), even for the uniform phases.

\subsection{Results}
\label{sec:iPEPS_results}
\begin{figure}[]
\begin{center}
\includegraphics[width=8.5cm]{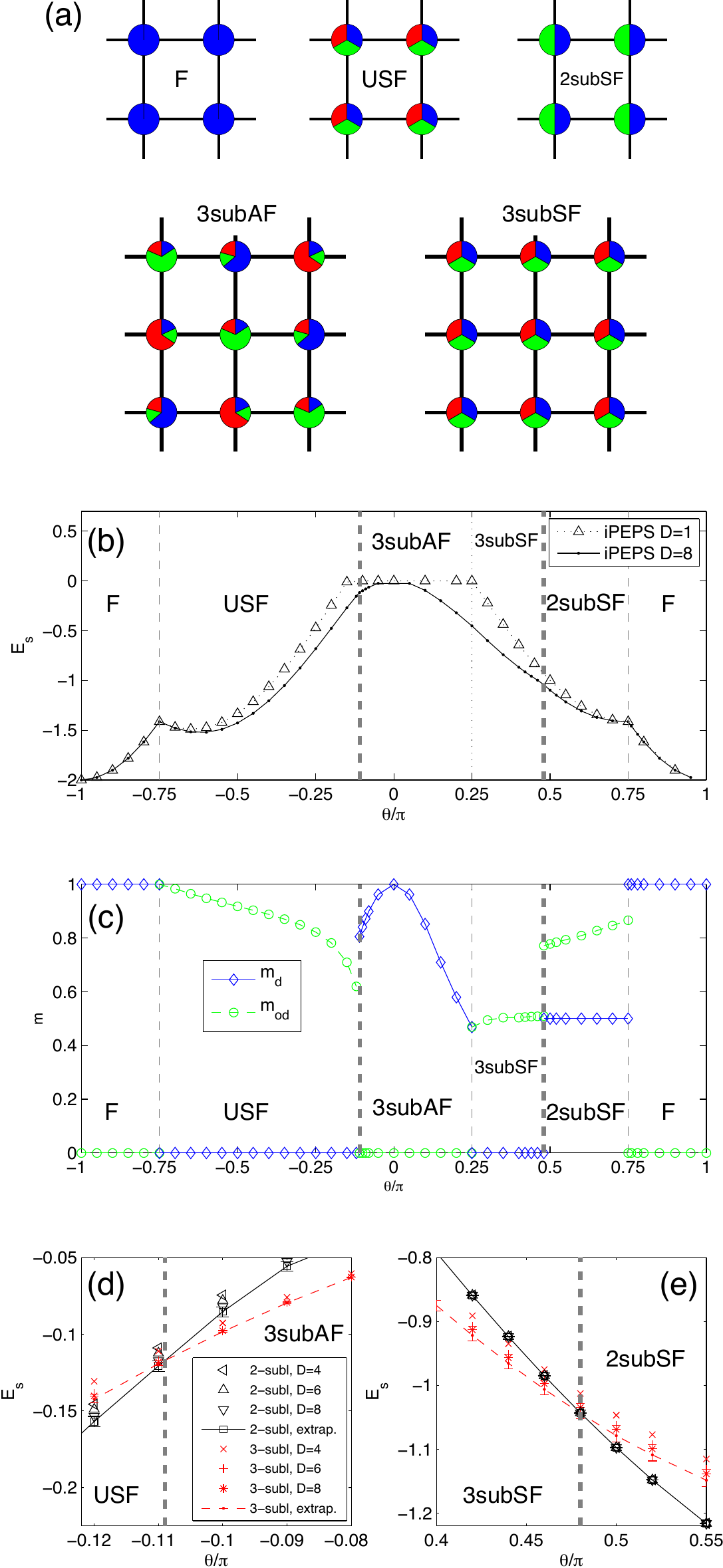}
\caption{iPEPS results as a function of $\theta$. (a)~Color densities in the iPEPS unit cell in the different phases.  (b)~Energy per site for different values of $D$. (c) Plot of the order parameters $m_{d}$ and $m_{od}$ obtained with $D=8$ which indicate long-range color order and/or (off-diagonal) superfluid order, respectively. (d)~First order phase transition between the USF and the 3subAF phase occurring for $\theta/\pi=-0.109(3)$. 
    (e)~First order phase transition between the 3subSF  and the 2subSF phase occurring for $\theta/\pi \approx 0.48 (1)$. 
    }
\label{fig:iPEPS}
\end{center}
\end{figure}

Our iPEPS results are summarized in Fig.~\ref{fig:iPEPS}: In Fig.~\ref{fig:iPEPS}(a) we present the local color densities of the three colors, given by pie-charts, on each site in the iPEPS unit cell for the different states in the phase diagram. Figure~\ref{fig:iPEPS}(b) shows the ground state energy obtained for different values of $D$ in the whole parameter range of $\theta$, and in Fig.~\ref{fig:iPEPS}(c) we plot the following order parameters,
\begin{eqnarray}
\label{eq:m}
 m_{d}&=&\sqrt{\frac{3}{2} \sum_{\alpha} \left| \langle \hat S^{\alpha \alpha}  - \frac{1}{3 } \rangle \right|^2}, \\
  m_{od}&=&\sqrt{\frac{3}{2} \sum_{\alpha \neq \beta} \left| \langle \hat S^{\alpha\beta}   \rangle \right|^2},
\end{eqnarray}
averaged over all sites in the unit cell. A finite value of $m_d$ indicates that there is long-range color (diagonal) order, whereas a finite $m_{od}$ implies off-diagonal long-range order (a superfluid phase).

In the following we discuss the results for the individual phases and then explain how we accurately determined the phase boundaries. 

In agreement with the findings in the previous sections, iPEPS predicts a ferromagnetic (F) phase between $-5\pi /4\le \theta \le -3\pi/4$. In this phase each site is occupied by the same color, as shown in Fig.\ref{fig:iPEPS}(a). Since the ferromagnetic state is a product state it can be trivially represented by an iPEPS, i.e. with a bond dimension $D=1$. This is why the energies  obtained with $D=1$ and the $D=8$ shown in Fig.\ref{fig:iPEPS}(b) are identical in this phase. 

In the range $-3\pi/4 \le \theta \le -0.109(3) \pi$ iPEPS confirms the uniform superfluid phase (USF). The states in this phase do not have color order, as can be seen in Fig.\ref{fig:iPEPS}(a) where all color densities on all sites are the same. However, we find a finite, uniform off-diagonal order parameter  $m_{od}$, indicating a superfluid phase.

In the range $-0.109(3)\pi \le \theta \le 0.48(1)\pi$ we clearly find the lowest variational energy with a 3-sublattice iPEPS ansatz. This range also includes the SU(3) symmetric point $\theta=\pi/4$, where iPEPS previously found a 3-sublattice N\'eel ordered state.\cite{SU3_Bauer} For $-0.109(3)\pi \le \theta <1/4 \pi$ we find a 3-sublattice diagonal order (3subAF in Fig.~\ref{fig:iPEPS}(a) with $\theta=0.2\pi$) where each site is occupied by one dominant color, and the three colors are arranged in a stripe pattern. The three sublattice stripe structure persists for $\theta>\pi/4$, however, in this range it is the offdiagonal terms in $m_{od}$ which are finite and the diagonal terms vanish (3subSF phase).
  
We also confirm the 2subSF phase with iPEPS in the range $0.48(1)\pi \le \theta \le 3/4 \pi$. Fig.~\ref{fig:iPEPS}(a) shows that the state has a uniform diagonal order (each site is equally populated by only two colors with the third color being absent), however, the off-diagonal terms exhibit a two-sublattice structure (not shown). 

Next we discuss the phase transition between the USF and 3subAF phase. 
We find clear evidence of a first order phase transition, with a jump of the order parameter at the transition value. In order to accurately determine the critical value $\theta_c$ we make use of the hysteresis effect in the vicinity of a first order phase transition: When we initialize a state in the USF phase and change $\theta$ to a value slightly above $\theta_c$ then the state will remain in the USF phase (because it is metastable). Similarly we can obtain the energy of the 3subAF state slightly below $\theta_c$. We  then determine $\theta_c=-0.109(3)$ from the intersection of the energies of the two states,  as shown in Fig.\ref{fig:iPEPS}(d). The error bar of the phase boundary includes the uncertainties when extrapolating the energies to the infinite $D$ limit.

In a similar way we have determined the location of the first order transition $\theta_c=0.48(1)\pi$ between the 3subSF and the 2subSF phase, as shown in Fig.~\ref{fig:iPEPS}(e). 

We did not find that the phase boundaries of the ferromagnetic phase change when going to larger bond dimensions. Thus, the location of the phase transitions is the same as predicted classically, i.e. $\theta=-3\pi/4$ and $\theta=3\pi/4$.

\section{Continuous time world line Monte-Carlo with cluster updates}
\label{sec:QMC}

  The iPEPS numerical simulations give access to informations at zero temperature. 
  Using quantum Monte Carlo (QMC) methods, finite temperature observables can be measured. 
  Moreover, QMC methods work in the full Hilbert space and thus is a non biased numerical method. 
  They give interesting complementary informations to the iPEPS and LFWT results. 
  
  Unfortunately, the so-called sign problem occurs as soon as some non-diagonal elements of the Hamiltonian are positive. 
  To study the model of Eq.~\eqref{eq:Ham}, we chose a basis of states consisting of the product states $\otimes_i|\alpha_i\rangle=|\alpha_0,\dots,\alpha_{N_s-1}\rangle$, where $\alpha_i$ is the color of the boson on site $i$ and $N_s$ is the number of sites of the lattice. 
  These are eigenstates of all the $\widehat S^{\alpha\alpha}_i$. 
  The sign of the non-diagonal elements of the Hamiltonian matrix is the sign of $J_\perp$. 
  The sign problem only allows us to simulate this model for $J_\perp<0$. 
  
  In some specific cases, the sign problem can be overcome. 
  On an infinite or open one dimensional chain, the physical properties of the $N$ color model are unchanged by the transformation $\theta\to-\theta$, whatever $N$. 
  This allows us to use the QMC method whatever the value of $\theta$, although mainly the SU($N$) symmetric point was considered.\cite{QMC_SU4_1D,entropy_SUN_chain}
  In two dimensions, the same transformation still conserves the physical properties for $N=2$ (see discussion in Sec.~\ref{sec:LFWT_results}), but not any more for $N>2$, where qualitatively different phases can be obtained for two opposite values of $\theta$, as can be seen in the phase diagram of Fig.~\ref{fig:product_states}. QMC is thus limited to $J_\perp<0$. 

\subsection{The algorithm}
\label{sec:QMC_alg}

  We use a continuous time world-line algorithm with cluster updates,\cite{Prokofev,CTQMC_loop}  adapted to our $3$-color model. 
  We express the partition function $Z$ as a path integral over the configurations $\phi:\tau\to\phi(\tau)$, where $\tau$ is the imaginary time going from 0 to $\beta=1/T$ ($T$ is the temperature) and $\phi(\tau)$ is a basis state. 
  The functions $\phi$ that contribute to $Z$ can be represented by $\phi(0)$ and by a set of world line crossings $\{(i,j,\tau)\}$ at which  the colors of two neighboring sites $i$ and $j$ at a time $\tau$ are exchanged. 
  A local configuration $c$ on a link $ij$ at time $\tau$ is represented by
  \begin{equation}
   c=\left(\begin{matrix}
     \alpha_i(\tau^+)\,\alpha_j(\tau^+) \\  \alpha_i(\tau^-)\,\alpha_j(\tau^-)
     \end{matrix}\right)
  \end{equation}
  Cluster algorithms are well known for 2-color (spin) models.
  Here we generalize them by chosing randomly two different colors $p$ and $q$ out of the $3$ and by constructing clusters on which only these two colors are encountered. 
  The steps to construct a cluster are the following. 
  We first randomly place the graphs depicted in the first column of Tab.~\ref{tab:cluster} on each link of the configuration using a Poisson distribution whose time constant $W(G)$ is given in the last columns of Tab.~\ref{tab:cluster} and accepting them if $\Delta_G(c)=1$ (if a color being neither $p$ nor $q$ appears in the local configuration, the graph is rejected). 
  Then we assign graphs to the world-line crossings between $p$ and $q$ colors with a probability proportional to $W(G)$. 
  At the places where no graph has been attributed, we follow the path with the same color. 
  Finally we follow each constructed cluster and exchange $p$ and $q$ on it with a probability $1/2$. 
  This completes one Monte Carlo step. 
  During the whole simulation, $n$ steps are performed. 
  
  \begin{table}
    \begin{displaymath}
      \begin{array}{!{\vrule width 1pt}c!{\vrule width 1pt}c|c|c!{\vrule width 1pt}c|c!{\vrule width 1pt}}
        \noalign{\hrule height 1pt}
        &\multicolumn{3}{c!{\vrule width 1pt}}{\Delta_G(c)}&\multicolumn{2}{c!{\vrule width 1pt}}{W(G)}\\
        G& \left(\begin{matrix}
        p\, p \\
        p\, p
        \end{matrix}\right)&
        \left(\begin{matrix}
        p\,q \\
        p\,q
        \end{matrix}\right)&
        \left(\begin{matrix}
        q\,p \\
        p\,q
        \end{matrix}\right)&
        \textrm{for }|J_\perp|\leq J_\parallel&
        \textrm{for }|J_\perp|\geq J_\parallel\\
       \noalign{\hrule height 1pt}
       \includegraphics[width=.025\textwidth]{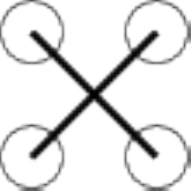}  &1 &- &1 &\epsilon& \frac{|J_\perp|}2-\frac{J_\parallel}2+\epsilon\\
       \includegraphics[width=.025\textwidth]{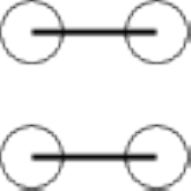}  &- &1 &1 &|J_\perp|-\epsilon&\frac{|J_\perp|}2+\frac{J_\parallel}2-\epsilon\\
       \includegraphics[width=.025\textwidth]{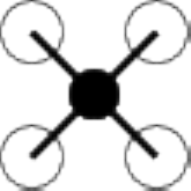}  &0 &1 &0 &J_\parallel-|J_\perp|+2\epsilon&2\epsilon\\
       \noalign{\hrule height 1pt}
      \end{array}
    \end{displaymath}
    \caption{
      Fabrication rules for clusters. 
      First column: all possible graphs $G$ that can be placed on a bond of the lattice at an imaginary time. 
      Three following columns: $\Delta_G(c)$, which is equal to 1 when $G$ is accepted and 0 otherwise. A dash ($-$)
      means that the graph is incompatible with the local configuration $c$. 
      Last two columns: time constant used to determine when a graph is inserted on a bond with a Poisson law. 
      $\epsilon$ is a small number ensuring the ergodicity ($\epsilon$ was set to $0.01$ in our simulations). 
      \label{tab:cluster}
    }
  \end{table}
  
  We use periodic square lattices of linear size $L$. 
  The goal is now to calculate averages of observables such as the correlations over the $n$ sampled configurations. 
  
  We define the diagonal correlation on site $j$ as
  \begin{equation}
  C(j)=\left\langle \sum_\alpha \widehat S^{\alpha\alpha}_0 \widehat S^{\alpha\alpha}_j\right\rangle -\frac1{N},
  \end{equation}
  and the associated structure factor
  \begin{equation}
  \tilde C(\mathbf k)=\frac1{4\pi^2} \frac{N}{N-1}\sum_jC(j)e^{i\mathbf k\cdot\mathbf r_j}, 
  \end{equation}
  with $\mathbf r_j$ the position vector of site $j$.
  This structure factor is normalized in such a way that $\frac{4\pi^2}{L^2}\sum_\mathbf k  \tilde C(\mathbf k)=1$.
  
  \subsection{Results}
  \label{sec:QMC_results}
  
\begin{figure}
    \begin{center}
      \includegraphics[width=0.3\textwidth]{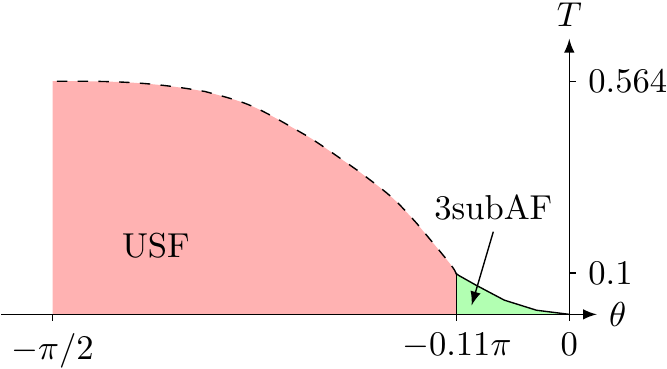}
      \end{center}
      \caption{Phase diagram for $-\pi/2<\theta<0$ obtained from QMC. 
      The dashed line is a KT phase transition, whereas the continuous line is a first order transition. 
      \label{fig:phase_diag}}
    \end{figure}

  We have studied the sector $-\pi/2<\theta<0$ of Fig.~\ref{fig:product_states} using the previously detailed algorithm. 
  Confirming LFWT and iPEPS results, we found two different phases at low temperature: the USF and the 3subAF stripe order. 
  We studied the thermal phase transitions generated by these two phases (see Fig.~\ref{fig:phase_diag}). 
  For USF, it is the KT type phase transition linked to the winding number $W$ described in Sec.~\ref{sec:transitions}. 
  The 3subAF stripe order gives rise to a first order transition due to the $P_3\times \mathbb Z_2$ symmetry breaking. 
  
\begin{figure}
      \begin{center}
        \subfigure[\label{fig:correlations_1}\,$\theta=-0.113\pi$, $T=0.075$]{\includegraphics[width=0.156\textwidth,trim=30mm 10mm 15mm 15mm,clip]{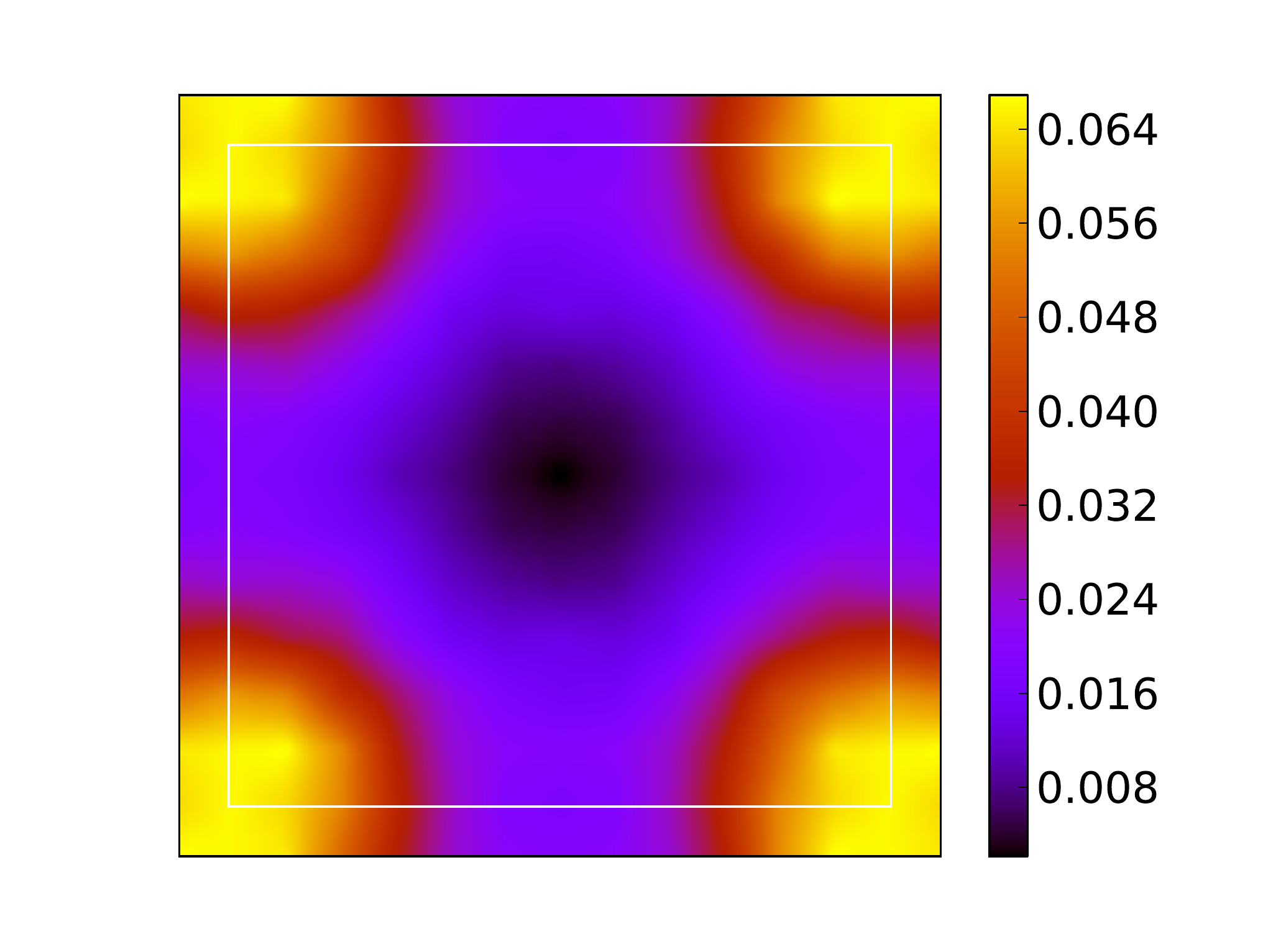}}
        \subfigure[\label{fig:correlations_2}\,$\theta=-0.113\pi$, $T=0.188$]{\includegraphics[width=0.156\textwidth,trim=30mm 10mm 15mm 15mm,clip]{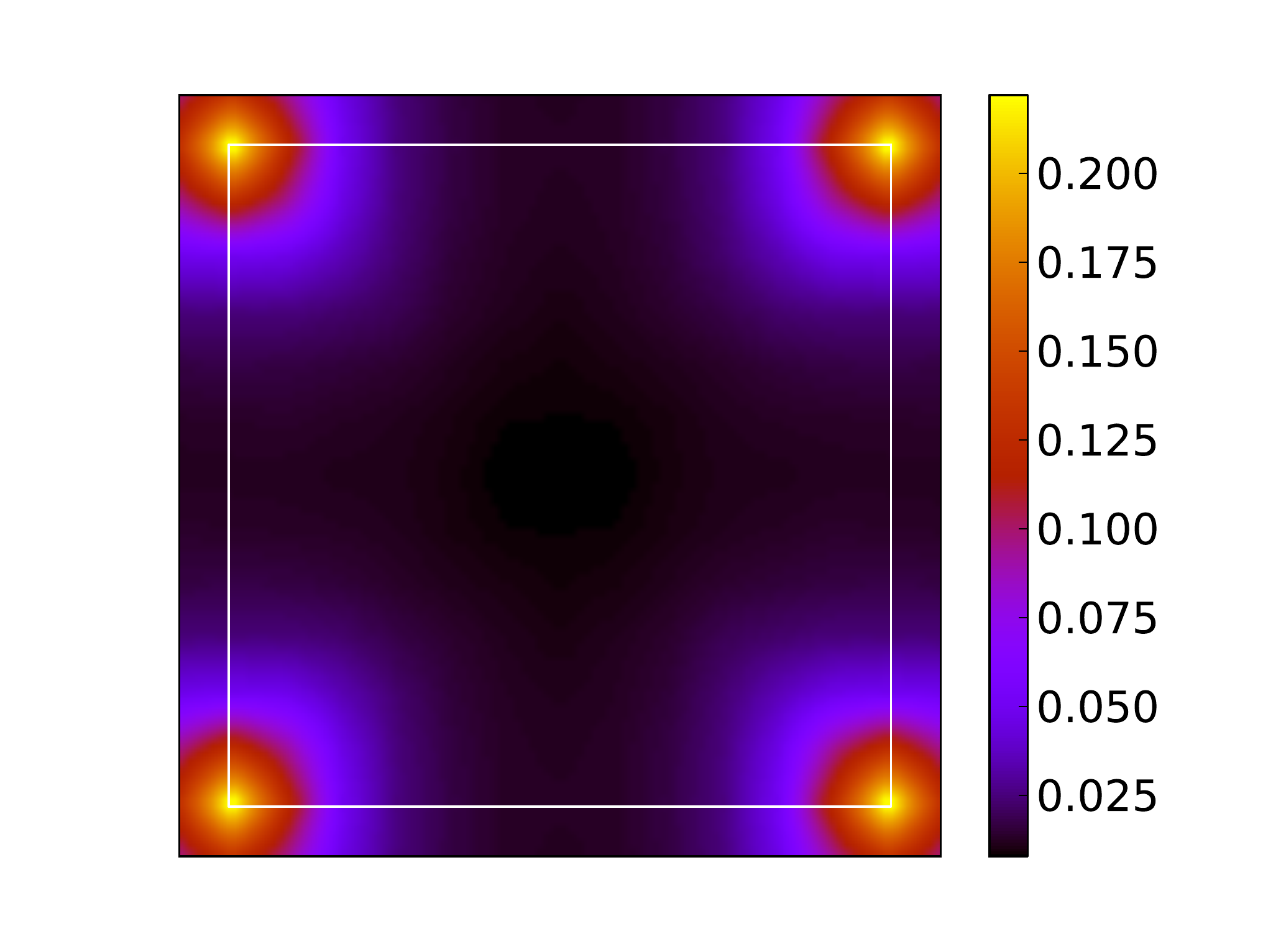}}
        \subfigure[\label{fig:correlations_3}\,$\theta=-0.101\pi$, $T=0.081$]{\includegraphics[width=0.156\textwidth,trim=30mm 10mm 15mm 15mm,clip]{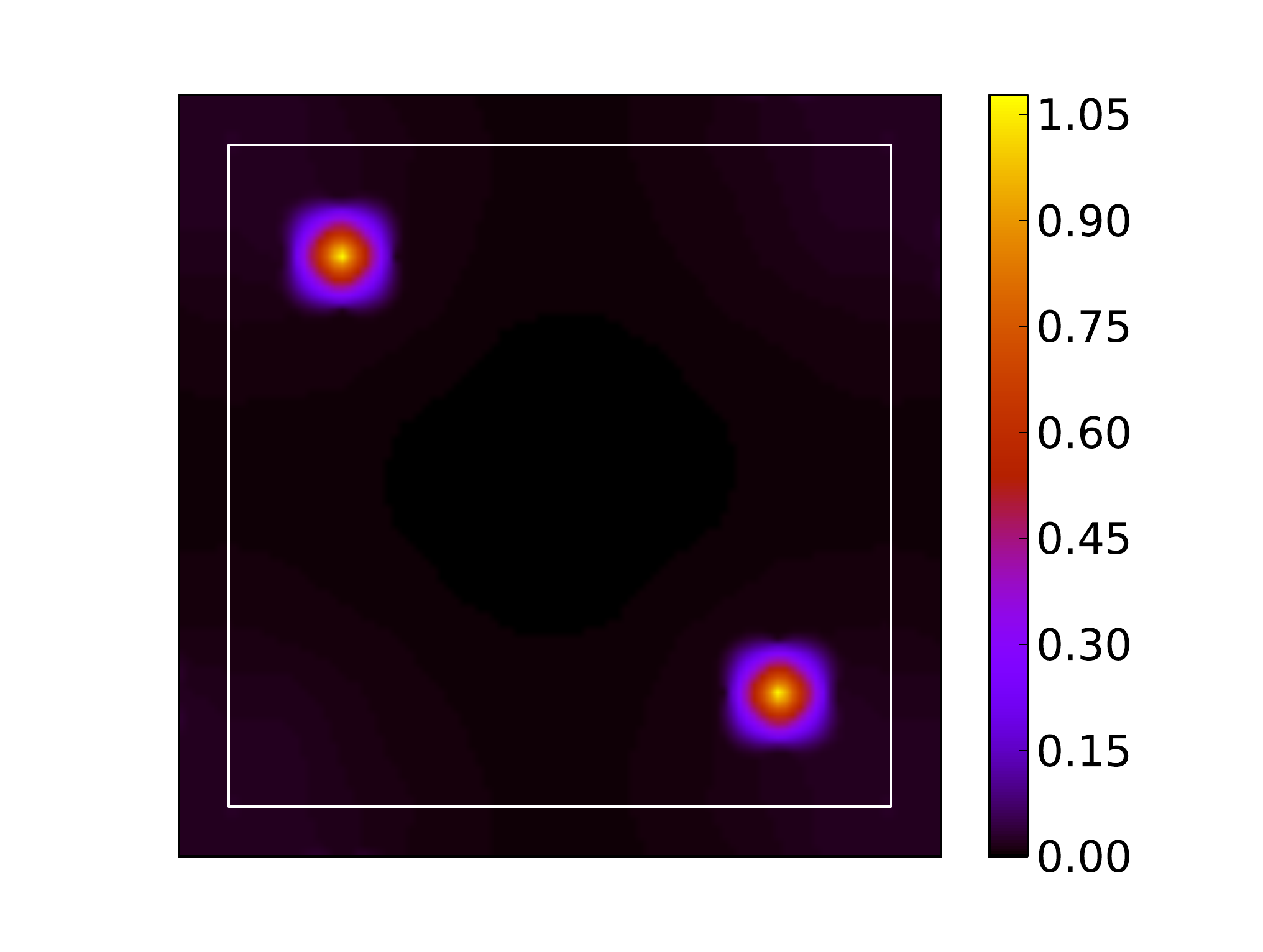}}
      \end{center}
      \caption{Structure factors obtained on a $L=12$ square lattice for the model of Eq.~\eqref{eq:Ham} with $N=3$. 
      The white square is the Brillouin zone.
      \label{fig:structure_factors}}
    \end{figure}

  The clearest sign of the $P_3\times \mathbb Z_2$ symmetry breaking is the structure factor in the reciprocal space. 
  Bragg peaks appear at two points of coordinates $\mathbf q=\pm( 2\pi/3,2\pi/3)$ or $\pm( 2\pi/3,-2\pi/3)$ depending on the direction of the stripes (see Fig.~\ref{fig:correlations_3}). 
  Their evolution with the size of the lattice as $L^2$ indicates long range order. 
  The transition temperature increases with more negative $\theta$'s (see Fig.\ref{fig:phase_diag}) and corresponds to a first order transition: coexistence of ordered and unordered phases occurs during the simulations. 
  When $\theta\simeq -0.11\pi$, no trace of the 3subAF phase is found at any temperature. 
  At large temperature, the structure factors have peaked maxima at point $\mathbf q=(\pi,\pi)$ (similar to Fig.~\ref{fig:correlations_2}), but the finite size effects indicate no long range order associated to these wave vectors. 
  
\begin{figure}
    \begin{center}
      \includegraphics[width=0.45\textwidth]{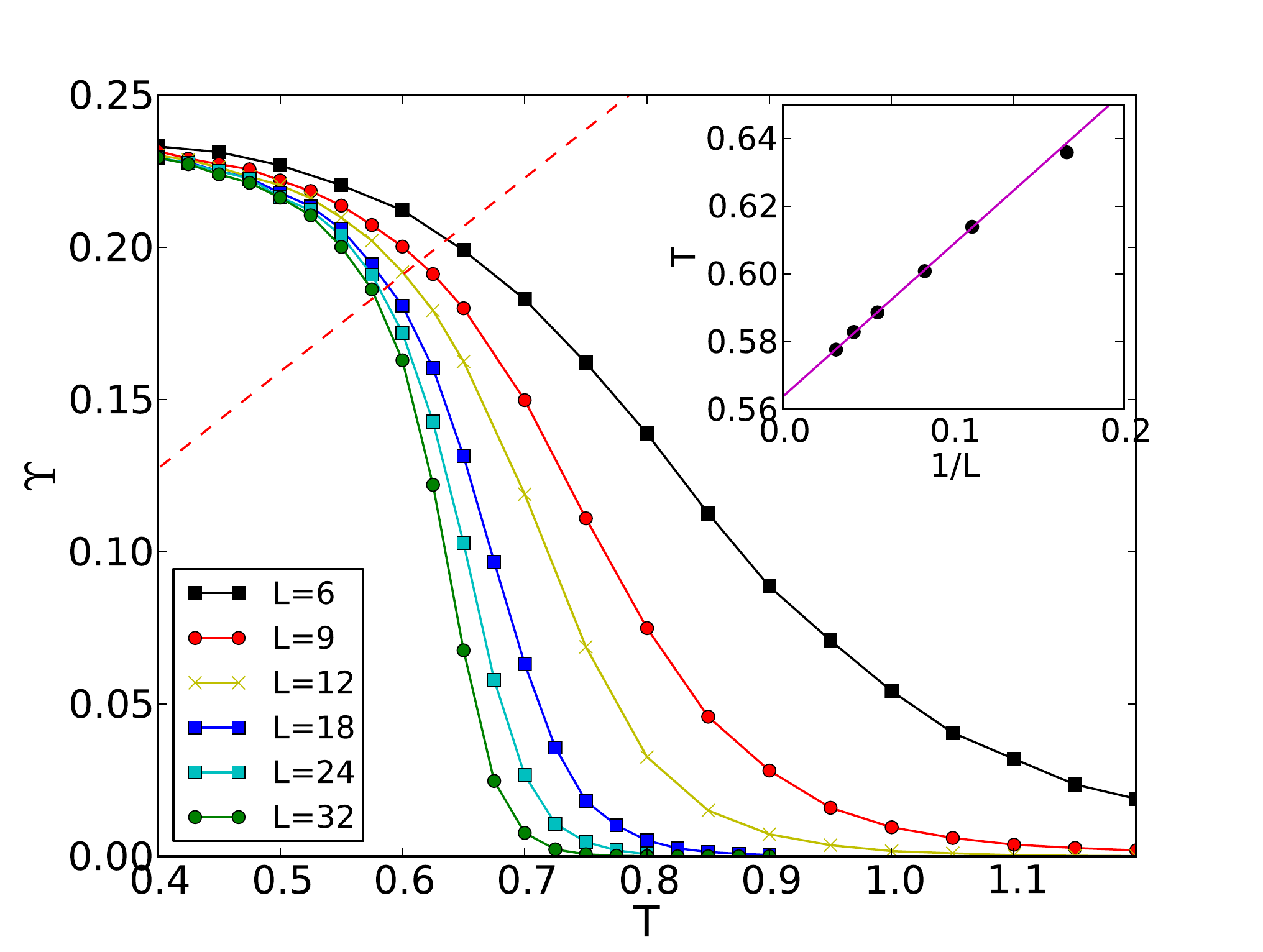}
      \end{center}
      \caption{Stiffness $\Upsilon$ versus temperature for different linear lattice sizes $L$ for $\theta=-\pi/2$ (precision of $0.001$: smaller than the symbol size). 
      The temperature of crossing with the dashed red line $\Upsilon=T/\pi$ is shown in the inset with the interpolation giving a critical temperature of $T_{KT}=0.564(2)$.
      \label{fig:W3}}
    \end{figure}

  For the KT type phase transition of the USF phase, we looked at the evolution of the stiffness $\Upsilon=\langle W^2\rangle/2T$ with the temperature.\cite{Winding_derivation} 
  The critical temperature is deduced from the intersection of the curves with the line $T/\pi$,\cite{universal_jump,winding_KT} extrapolated to the thermodynamic limit.  
  At $\theta=-\pi/2$ it is $T_{KT}= 0.564(2)$ (see Fig.~\ref{fig:W3}).
  Even if this phase cannot be characterized by the structure factor as there is no long range order, it looks quite different from the high temperature phase. 
  The peak at $\mathbf q=(\pi,\pi)$ is much broader, as illustrated on Fig.~\ref{fig:correlations_1}. 
  
  \begin{figure}
    \begin{center}
        \includegraphics[width=0.45\textwidth]{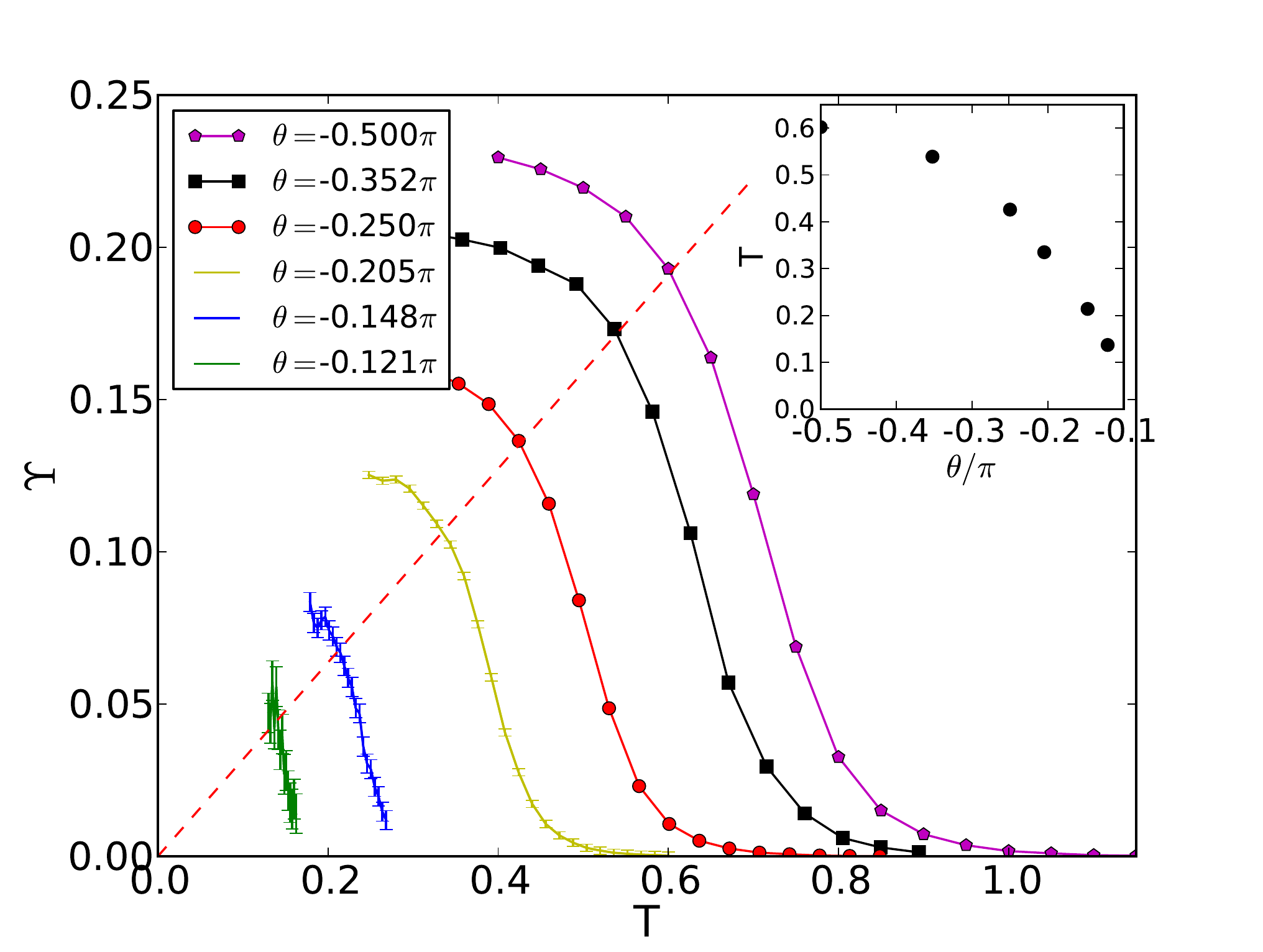}
      \end{center}
      \caption{Stiffness $\Upsilon$ versus temperature for different $\theta$ and a lattice linear size of $L=12$. 
      The temperature of crossing with the dashed red line $\Upsilon=T/\pi$ is shown in the inset and is compatible with a critical temperature reaching zero for $\theta\simeq 0.11\pi$. When not visible, the error bars are smaller than the symbol size. 
      \label{fig:W2}}
    \end{figure}

\section{Results and conclusions}
\label{sec:CCL}
We have studied a quantum version of the $N=3$-state Potts model on the square lattice which can be seen as a generalization of the well known $S=1/2$ XXZ model. 
We have presented the full phase diagram as a function of the coupling parameter $\theta$, where $J_\parallel=\cos\theta$ and  $J_\perp=\sin\theta$ are the exchange amplitudes between respectively identical and different colors. 
The model corresponds to the infinite repulsion limit of the fermionic Hubbard model in the $\theta=\pi/4$ case, and of the bosonic
Hubbard model for $-\pi<\theta<-\pi/4$. 

Three complementary methods have been used that give coherent results: linear flavor wave theory, iPEPS and quantum Monte Carlo. 
The resulting ground state phase diagram as a function of $\theta$ is given in Fig.~\ref{fig:product_states} and includes 5 phases: three superfluid phases (USF, 2subSF, and 3subSF), a ferromagnetic phase (F), and a three-sublattice color-ordered phase (3subAF). 
The temperature dependent properties have been determined for $-\pi/2<\theta<0$ and are presented in the phase diagram of Fig.~\ref{fig:phase_diag}. 
The 3subAF phase gives rise to a first order phase transition and the USF phase to a KT type phase transition. 

The phase diagram for $N=3$ has many similarities with the one from the $S=1/2$ XXZ model ($N=2$): A uniform superfluid and a $N$-sublattice superfluid characterized by $N-1$ phase angles, a ferromagnetic phase that can take $N$ different colors, and a $N$-sublattice antiferromagnetic, stripe-ordered phase. Each of these phases for $N=3$ can be seen as a generalization of the corresponding phase for $N=2$. 
However, the $N=3$ case has one additional phase: the 2subSF phase, which is not strictly speaking a generalization of a phase of the $N=2$ case, but is in fact equivalent to the 2-sublattice superfluid phase of the $N=2$ case (since only two colors are present in this phase). 
Thus it is interesting that for $N=3$ a 2-color superfluid is competing with a 3-color superfluid, and that both phases are stable in a certain parameter range. 
One finds also a competition between a 2-sublattice and a 3-sublattice color ordered state (which are degenerate at the classical level). 
In this case, however, it is the 3-sublattice color-ordered state which has always a lower energy than the 2-sublattice state.

Finally, coming back to possible experimental realizations, three phases of this phase diagram should be accessible with cold
atoms loaded in optical lattices: the USF phase and the F phase with bosonic particles, and the 3subAF phase at the symmetric AF point with fermionic particles. 

\subsection{Acknowledgment}
We thank S. E. Korshunov for useful discussions about the superfluid transitions, and the Swiss National Fund for financial support.

\bibliographystyle{apsrev4-1}
\bibliography{refs_article_SU3.bib}

\end{document}